# New data, new possibilities: Exploring the insides of Altmetric.com

Nicolás Robinson-García[1], Daniel Torres-Salinas[2], Zohreh Zahedi[3] and Rodrigo Costas[3]

[1] EC3: Evaluación de la Ciencia y de la Comunicación Científica, Departamento de Información y Documentación, Universidad de Granada, Spain

[2] EC3Metrics, Granada, Spain

[3] Centre for Science and Technology Studies, Leiden University, The Netherlands

**Abstract**

This paper analyzes Altmetric.com, one of the most important altmetric data providers currently used. We have analyzed a set of publications with DOI number indexed in the Web of Science during the period 2011-2013 and collected their data with the Altmetric API. 19% of the original set of papers was retrieved from Altmetric.com including some altmetric data. We identified 16 different social media sources from which Altmetric.com retrieves data. However five of them cover 95.5% of the total set. Twitter (87.1%) and Mendeley (64.8%) have the highest coverage. We conclude that Altmetric.com is a transparent, rich and accurate tool for altmetric data. Nevertheless, there are still potential limitations on its exhaustiveness as well as on the selection of social media sources that need further research.

**Título: Nuevos datos, nuevas posibilidades: Revelando el interior de Altmetric.com**

**Resumen**

Este trabajo analiza Altmetric.com, una de las fuentes de datos altmétricos más usadas actualmente. Para ello hemos cruzado un set de publicaciones con DOI indexadas en la Web of Science para el periodo 2011-2013 con la API de Altmetric.com. Solo el 19% de las publicaciones de nuestro set estaban indexadas en Altmetric.com. Este recurso obtiene datos altmétricos de 16 redes sociales distintas. No obstante, cinco de ellas representan el 95.5% del set de datos recuperado. Twitter (87.1%) y Mendeley (64.8%) cubren un mayor número de publicaciones. Concluimos destacando Altmetric.com como una herramienta rica, transparente y precisa en sus datos altmétricos. No obstante, ofrece aún algunas dudas acerca de la exhaustividad de la recuperación así como de la selección de fuentes que requieren más investigación.

**Palabras clave:** Altmetric.com; Twitter; Mendeley; indicadores altmétricos; impacto social; cobertura; Web 2.0

Introduction

Citation analysis has been traditionally confronted with different and opposed views as to its suitability to quantitatively measure the 'scientific impact' of publications. In brief, these have to do with citation biases, publication delays or process biases derived from peer review limitations (**Bollen; van de Sompel**, 2006). Several alternatives have been proposed, especially since the 1990s and the expansion of the Internet and the digital media. Among others here we highlight the use of acknowledgments or *influmetrics* (**Cronin; Weaver**, 1995), web links or *webometrics* (**Almind; Ingwersen**, 1997) and usage metrics (**Kurz; Bollen**, 2010). However, the most recent proposal as an alternative to traditional citation analysis has become a hot topic within the bibliometric community. *Altmetrics* or the use of social media-based indicators to quantify the social impact of scholarly information was first proposed by **Priem** *et al.* (2010). Since then it has become a research front of itself producing its own scientific corpus as it has been received by the research community.

Altmetric proponents claim that such indicators have the potential to complement or improve the more traditional scientific evaluation systems (**Priem**, *et al*, 2010). They base their arguments stating that almetric indicators provide a wider picture of the relevance and impact of scientific contributions (or 'research products') (**Piwowar**, 2013); also, they are produced at greater speed than citations and end with the monopoly exerted by citation indexes as they come from open sources. However, their strongest claim is that they can capture other aspects of impact different from those derived from citation counting. However, the reality is that they are still under-developed and much study is needed before confirming such arguments, which are currently either questionable or simple promises (**Wouters; Costas**, 2012).

Hence, there are still serious concerns as to the meaning of these indicators (**Torres; Cabezas; Jiménez**, 2013; **Torres-Salinas; Cabezas-Clavijo**, 2013) and the suitability of the sources (**Thelwall** *et al.,* 2013). So far, studies have reported 1) a relatively weak correlation with citations (i.e., **Thelwall** *et al.,* 2013; **Costas; Zahedi; Wouters**, 2014), 2) their potential to offer complement aspects of impact remains unknown and 3) Twitter, blogs mentions, Mendeley readers, F1000 recommendations or news outlets seem to be among the most relevant sources (**Li; Thelwall**, 2012; **Li; Thelwall; Giustini**, 2012; **Haustein** *et al.,* 2013; **Costas; Zahedi; Wouters**, 2014; **Zahedi; Costas; Wouters**, in press). Regarding this latter issue, many tools have appeared in the last few years recollecting and providing these metrics. The main ones are ImpactStory.org[1], Plum Analytics[2] and Altmetric.com[3].

Altmetric.com is currently one of the most important altmetric data providers. It captures information regarding the impact of a paper from various social media sources developing a weighted score. In order to do so it disambiguates links to articles, unifying links to PubMed records, Arxiv identifiers, DOI numbers or publisher's sites. Although some have warned against the use of aggregated altmetric scores (**Davis**, 2013), there has been less debate about the richness and diversity of the data provided. One of the major problems potential users face when dealing with this source is that such diversity and richness of data is actually difficult to grasp. Although the web company provides extensive information of its contents (**http://support.altmetric.com**) one would still have difficulties in understanding the broadness of the data and possibilities that this source could provide.

The aim of this paper is to explore Altmetric.com as a source for developing altmetric indicators. In order to unveil the potential use of this tool, we provide a comprehensive and practical view on the contents available in Altmetric.com. Specifically, we will answer the following research questions:

1. Which data sources are included in Altmetric.com and how are they structured?

2. What is the coverage of Altmetric.com and which data sources cover more altmetric impact of publications?

For this we have performed a practical extraction of data from Altmetric.com and carried out a detailed analysis of the data provided by this tool.

**Material and methods**

In order to explore Altmetric.com, we selected all publications between 2011 and 2013[4] indexed in the Web of Science database using the CWTS (University of Leiden) in-house version. From this set of papers we selected only those which included a DOI number. In January 2014 we matched a total of 2,792,706 DOI numbers with the Altmetric API (**https://api.altmetric.com/**). We retrieved a total of 516,150 records from the Altmetric API. This means that roughly 19% of all publication with DOI number during the study time period had received some kind of social media attention. However, we most note that there are errors on some of the unique DOIs present in Altmetric.com. Also, not all papers in Altmetric.com include DOI information. For each record we obtained a file on Javascript Object Notation format (JSON)[5]. The JSON files include raw data collected by Altmetric.com for each publication. **Table 1** shows the structure of each file indicating the type of information provided for each section.

| Description | Example of fields extracted |
|---|---|
| Summary of metrics as shown in the Altmetric.com bookmarklet | "counts":{"readers":{"mendeley","citeulike","connotea"},"facebook":{"unique_users_count","unique_users":[],"posts_count"},"blogs":{"unique_users_count","unique_users":[],"posts_count"},"news":{"unique_users_count","unique_users":[],"posts_count"},"pinterest":{"unique_users_count","unique_users":[],"posts_count"},"reddit":{"unique_users_count","unique_users":[],"posts_count"},"twitter":{"unique_users_count","unique_users":[],"posts_count"},"video":{"unique_users_count","unique_users":[],"posts_count"}},"linkedin":{"unique_users_count","unique_users":[]","posts_count","total":[]"... |
| Bibliographic description of the paper | "citation":{"title","authors":[],"pubdate","volume","issue","startpage","endpage","doi","PMID","arxiv_id","journal","altmetric_jid","links":[],"first_seen_on"} |
| Comparison and evolution of the aggregated Altmetric score | "altmetric_score":{"score","score_history":{"1d","2d","3d","4d","5d","6d","1w","1m","3m","6m","1y","at"},"context_for_score":{"all":{"rank","mean","median","sample_size","sparkline","total_number_of_other_articles","this_scored_higher_than","this_scored_higher_than_pct","percentile","rank_type":"approximate"},"similar_age_3m":{"rank","mean","median","sample_size","sparkline","total_number_of_other_articles","this_scored_higher_than","this_scored_higher_than_pct","percentile","rank_type":"approximate"},... |
| Demographics (Twitter): Public type and country | "demographics":{"poster_types":{"member_of_the_public","researcher","practitioner","science_communicator"},"geo":{"twitter":{"*Country*":"*number of users*"}}} |
| Altmetric data disaggregated by provider | "posts":{"twitter":[{{"url","posted_on","license","summary","author":{"name","image","id_on_source","followers"},"tweet_id"}}],"blogs":[{"title"{"title","url","posted_on","summary","author":{"name","url","description"}}],"facebook":[{"title","url","posted_on","summary","author":{"name","url","facebook_wall_name","image"","id_on_source"}},{"url","posted_on","summary","author":{"name","url","facebook_wall_name","image","id_on_source"}}],"googleplus":[{{"title","url","posted_on","summary","author":{"name","url","image","id_on_source"}}],... |

**Table 1. Disaggregated structure from a record provided by the Altmetric API**

As observed, five distinctive parts were identified. The first section is a summary with the global scores by source from which counts have been retrieved. Secondly, a brief description of the scientific paper is given including not only the bibliographic reference but also information such as the date when the paper was first included in the system or alternative links to the paper. The third part of the file offers a temporal evolution of the aggregated altmetric score for different time periods, along with comparisons with the journal's scores. Forth, a demographic display is shown by country and public type. This information is based on the Twitter account of users mentioning the paper. Finally, the last section includes a display with all the information and fields recorded in the system derived from each of the sources from which Altmetric.com retrieves the data.

**Description of sources collected by Altmetric.com**

16 sources were identified in Altmetric.com. In **table 2** we display each source including a brief description, the type of metric they measure and the data fields retrieved by Altmetric.com. Each record keeps a historical track of all metrics recorded since 2011 or since the inclusion of the paper in the system. In order to capture this data, Altmetric.com identifies mentions through link recognition. The only exception is done with blogs and news, where they also employ a tracker mechanism using text-mining techniques in order to capture those mentions which do not link to the publication. Such techniques are employed only for English language sources.

As observed, the most common type of metrics collected are discussions and mentions (four sources for each metric), followed by readership counts (Mendeley, Connotea and Citeulike). Then, other similar metrics to these can be seen such as videos, reviews or 'Question and Answer' discussion threads. As observed, with the exception of Research Highlights, which includes citation data retrieved from the highlights section of Nature magazine, all sources are of a 2.0 nature. Also, some of these sources may be biased towards certain fields. For instance, F1000 is a post-publication peer review service of Biomedical and Medicine research (**Waltman; Costas** 2014). Also, Stack Exchange is especially used by researchers from Computer and Natural Sciences.

| Source | Description | Type of metrics | Data elements |
|---|---|---|---|
| *Blogs* | Manually-curated RSS list | Discussion | Blog title; post title; post URL; publication date and time; summary; author name; author URL; author description |
| *News* | Manually-curated RSS list | Discussion | News title; news URL; publication date and time; license; summary; news media name; news media URL; news media id; news media image |
| *Reddit* | News provider | Discussion | News title; reddit URL; publication date and time; author name; author URL; author id; followers; subreddit |
| *Facebook* | Social network | Mentions | Mention title; URL mention; publication date and time; summary; author name; author URL; Facebook wall name; author image; author id |
| *Google Plus* | Social network | Mentions | Mention title; URL mention; publication date and time; summary; author name; author URL; author image; author id |
| *Pinterest* | Social network | Mentions | Mention URL; mention image; publication date and time; summary; author name; pinboard |
| *Twitter* | Microblogging | Mentions | URL; publication date and time; license; summary; author name; author image; number of followers, tweet id; type of public; country |
| *Stack Exchange* | Question & Answer site | Discussion | Thread title; thread URL; publication date and time; summary; author id |
| *Citeulike* | Social bookmarking | Readers | Total count of bookmarks |
| *Connotea* | Social bookmarking (discontinued) | Readers | Total count of bookmarks |
| *Mendeley* | Social bookmarking | Readers | Total count of bookmarks |
| *F1000* | Pospublication peer review service | Reviews | Recommended in F1000; publication date (probably of the last update); type of recommendation |
| *YouTube* | Video sharing site | Video | Video title; video URL; video image; publication date and time; license; summary; embed type; YouTube id; author name; author id |

| Source | Description | Type of metrics | Data elements |
|---:|---|---|---|
| *LinkedIn Groups* | Professional social network | Mentions | Total unique users; unique users name; total posts; post title; summary; publication date and time; author name; author description; post URL; group logo URL; group name; group description |
| *Research Highlights* | Nature highlights | Citations | Highlight URL; date added to Altmetric.com; highlight title; total highlights; bibliographic description of highlight; first seen |
| *Misc* | Others | Others | This field includes data from different social media sources which are added on authors' request (**Adie**, 2014) |

**Table 2. Summary of data elements provided by Altmetric.com by data sources**

With the exception of the *Misc* field which is devoted to other media sources not included in the original set of Altmetric.com, all are included when calculating the aggregated Altmetric score of each paper. Most of this information can be displayed through the Altmetric.com bookmarklet (**Figure 1**). However, some differences have been noted between the records retrieved from the Altmetric API and those displayed in the Altmetric bookmarklet: some indicators and data elements are not displayed in the breakup of the bookmarklet (e.g. all tweets and retweets) or discrepancies between the information provided between the sources (e.g. occasional errors in the Q&A threads).

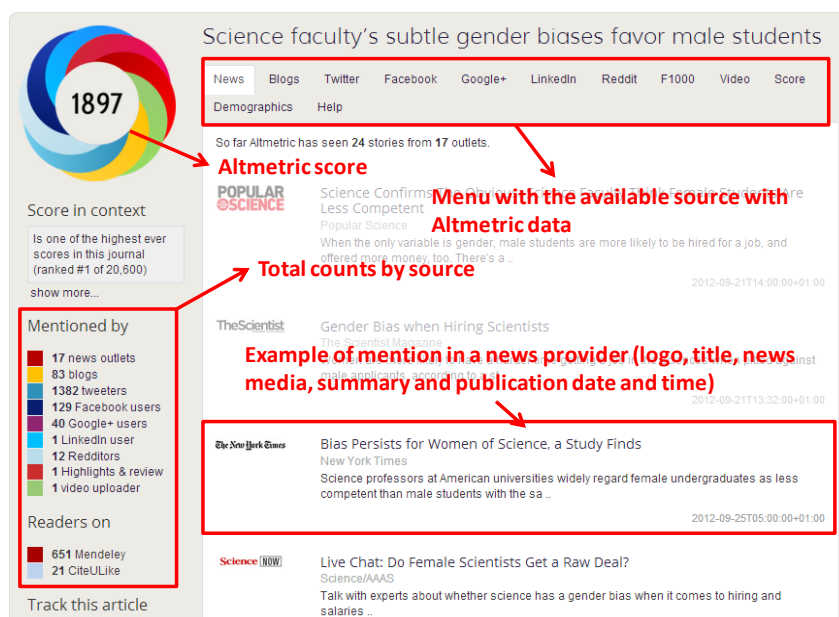

**Figure 1. Example of data provided by the Altmetric.com bookmarklet**

### Coverage of Altmetric.com for WoS publications with DOI in 2011-2013[4]

From the total of publications in the original sample, only 19% were included in Altmetric.com reporting some type of altmetric impact (**Figure 2**). Twitter is the source providing more altmetric data (87.1%) followed by Mendeley (64.8%). None of the other social media reaches values higher than 20% of the total share of papers with altmetric indicators associated, although Facebook reaches a total share of 19.9% of papers included in Altmetric.com.

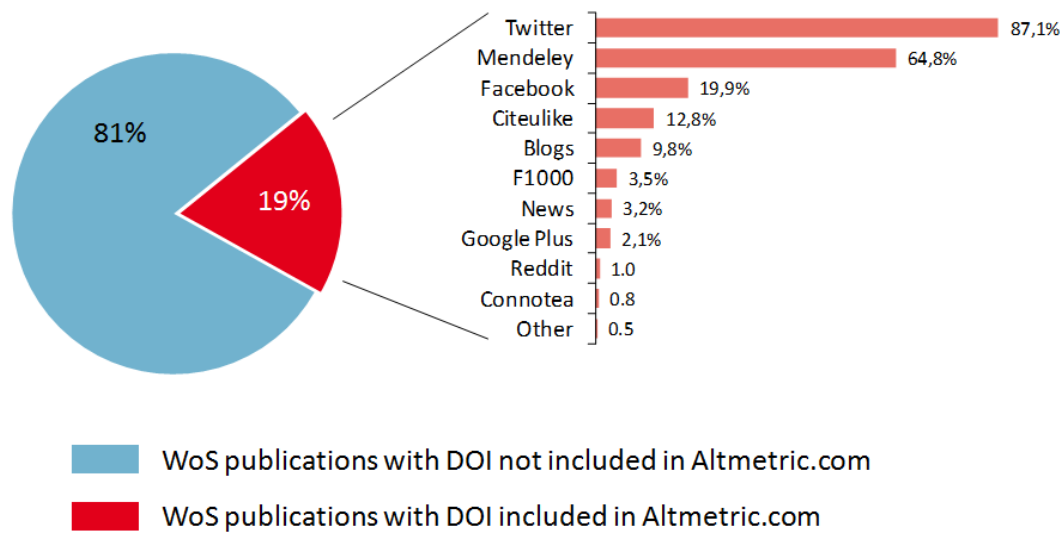

**Figure 3. Coverage of WoS papers in Altmetric.com by social media for the period 2011-2103[4]**

In **table 3** we include further information on the number of papers including metrics, total counts of each metric and unique users for the five top sources (Twitter, Mendeley, Facebook, Citeulike and blogs). These sources are present in 95.5% of the total share of papers retrieved from Altmetric.com. Although Twitter is the social media with the most mentions, Mendeley includes a higher number of users bookmarking scientific papers. These two data sources are the most expanded social media among all the altmetric sources analyzed. Indeed, the presence of mentions to scientific papers from social media such as Facebook, Citeulike or even blogs, never reaches 5% of the total papers with DOI indexed in the Web of Science during the studied time period.

| Social media | Papers | Total counts | Unique users | % Papers in WoS |
|---:|---:|---:|---:|---:|
| Twitter | 449,493 | 1,819,194 | 1,621,396 | 16.1 |
| Mendeley | 334,616 | 2,631,396 | 2,631,396 | 12.0 |
| Facebook | 102,923 | 197,449 | 182,422 | 3.7 |
| Citeulike | 65,799 | 130,756 | 130,756 | 2.4 |
| Blogs | 50,529 | 84,927 | 75,946 | 1.8 |

**Table 3. Coverage of Altmetric.com by social media to papers indexed in Web of Science for the 2011-2013[4] time period**

**Discussion and concluding remarks**

In this paper we analyzed Altmetric.com as an altmetric data provider for analyzing the altmetric impact of scientific publications. The main issue this type of sources have is the difficulties that entail identifying mentions to scientific papers, similarly to the shortcomings found when using webometric techniques (**Thelwall**, 2011). Although Altmetric.com states that they do serious efforts on link disambiguation (**http://support.altmetric.com**), there is still an important lack of research on the exhaustiveness, precision and correctness of the information retrieved by these tools (e.g. How many mentions is Altmetric.com missing from the covered sources?). This is specially relevant when analyzing the retrieval method for identifying mentions to scientific papers in more problematic sources such as blogs or news media. Here, a tracker mechanism based on text-mining techniques is applied as a complement to the link recognition method. However, it is applied to a manually-curated list of resources, not being evident the criteria followed for selecting them (**http://www.altmetric.com/sources-blogs.php**). Also, this technique is applied only for English language sources while for non-English sources only direct links to publications are considered (**http://www.altmetric.com/sources-news.php**), which inserts an important language bias that needs to be considered when studying publications from different languages.

Conceptually speaking, a very serious limitation is related to the sources covered by Altmetric.com. The reasons why these and no other sources are covered is a relevant question. Particularly in an environment of increasingly growing social media tools. In fact, this shortcoming applies to all altmetric providers as they do not always empirically or conceptually justify their selected sources. As such, one could argue that if Facebook is included, why not the Spanish Tuenti? If Twitter is covered, why not Tumblr, or the Spanish 'Menéame' along with Reddit? In the same line, related with scientific research it is worth mentioning the omission of scientific social networks such as Academia.edu or ResearchGate which seem to be used by many researchers (**Mas-Bleda; Thelwall; Kousha; Aguillo**, 2014). In this sense, some improvements have been reported, and on April 7, 2014, Altmetric.com reported the inclusion of the Chinese Weibo as a new source (**Adie**, 2014).

Probably, the reason for the selection of the current sources is more practical than conceptual (these sources are popular, have public APIs, are international, etc.) and although with limitations, finding and scanning mentions to research outputs across them is relatively feasible. However, technical issues should not avoid a more conceptual and theoretical discussion on what should be covered and the possible limitations or biases of the current sources, similarly to the analyses on coverage and limitations of other bibliometric databases such as the Web of Science, Scopus or Google Scholar (e.g. **Jacso**, 2009).

Our results show that from the 16 sources covered by Altmetric.com only 5 represent 95.5% of the total share of publications with altmetrics. This opens the question of the relevance of the sources and whether the smaller ones can really provide a meaningful evidence of impact. Indeed such concentration in a small number of social media has already been discussed elsewhere (**Priem** *et al.*, 2012; **Cabezas-Clavijo; Torres-Salinas**, 2010). The most important sources are Twitter and Mendeley (**Figure 2**). These sources are the ones that seem more promising for determining the type of impact altmetric data provide, as they show a higher density and therefore more reliable metrics could be extracted from them. As observed in our results, while Twitter seems to show data related to a larger number of publications, Mendeley shows higher figures (**Table 2**), including a larger number of counts and users. In this sense, this latter tool seems to have expanded much among the scientific community (**Haustein** *et al.*, 2014). Surprisingly, Altmetric.com does not collect readership data (i.e., Mendeley data) unless other bibliometric indicators are collected (**Costas; Zohedi; Wouters**, 2014).

All in all, Altmetric.com is indeed a very relevant open tool and data provider, which shows high quality and transparent data related to mentions in social media to scientific publications. The recent partnership established between ImpactStory (another important altmetric tool) and Altmetric.com (**Piwowar**, 2014) is a clear recognition of the value of this tool. Our study highlights the richness of the data collected. This richness is reflected in the fact that not only metrics about the counts and mentions on the different social media tools are recorded, but also data elements about their users and their origin or the dates of their mentions, for instance. As it stands, this data collection has two important positive implications. First, the fact that the data are stored and recorded permanently allows the reproducibility of the results and retrospective analysis, thus giving a solution to the problem of volatility of altmetric data (**Wouters; Costas**, 2012). Secondly, the abundance of data elements recorded opens the possibilities for further analyses that go beyond the simple counting of mentions. For example, the possibility of analyzing types of audience, the interests of these audiences, their relationships, etc. are new possibilities not yet explored.

Finally, our study shows that there are still important issues that need to be resolved to fully understand altmetric data. Our results indicate that more research is needed for understanding the methodologies for retrieving valid and reliable altmetric data. In the same line, the selection of social media sources must be rigorous and critical, attending to its use within the different communities and audiences and avoiding potential discipline or language biases.

**Acknowledgments**

The authors would like to thank Erik van Wijk from CWTS for helping in the retrieval of the data. Euan Adie from Altmetric.com clarified some of our concerns on the data. Stefanie Haustein contributed with her comments which improved the final version of the manuscript. Nicolás Robinson-García is currently supported with a FPU grant from the Spanish Ministerio de Economía y Competitividad.

**Notes**

[1] http://impactstory.org. Founded by Jason Priem and Heather Piwowar in 2011, it was originally called Total-Impact.

[2] http://www.plumanalytics.com/. Founded in late 2011 by Andrea Michalek and Mike Buschman, it has recently been acquired by EBSCO Publishing.

[3] http://www.altmetric.com/. Founded by Euan Audie in 2011, it has become one of the main altmetric providers.

[4] The publication year 2013 is not complete. Only one third of the publications were uploaded in the system at that time. In any case, this is not problematic for our analysis as we are just doing a descriptive analysis of the presence of Altmetric.com covered mentions across available scientific publications.

[5] For more information about the JSON format the reader is referred to http://en.wikipedia.org/wiki/JSON

Paper published in *El profesional de la información*, vol. 23, n.4, pp. 359-366
doi:10-3145/epi.2014.jul.03**References**

**Adie, E.** (2014). "Announcing Sina Weibo support". *http://www.altmetric.com/blog/announcing-sina-weibo-support/*

**Adie, E.** (2014) "Personal communication".

**Almind, T.; Ingwersen, P.** (1997). "Informetric analyses on the world wide web: Methodological approaches to 'webometrics'". *Journal of Documentation,* vol. 53, n. 4, pp. 404-426. *http://dx.doi.org/10.1108/EUM0000000007205*

**Altmetric.com.** "Knowledge Base". *http://support.altmetric.com/knowledgebase*

**Bollen, J.; Van de Sompel, H.** (2006). "Mapping the structure of science through usage". *Scientometrics*, vol. 69, n. 2, pp. 227-258. *http://dx.doi.org/10.1007/s11192-006-0151-8*

**Cabezas-Clavijo, Á.; Torres-Salinas, D.** (2010). "Los investigadores en la ciencia 2.0: El caso de PLOS One". *El profesional de la información*, vol. 19, n. 4, 431-434.

**Costas, R.; Zahedi, Z.; Wouters, P.** (2014). "Do 'altmetrics' correlate with citations? Extensive comparison of altmetric indicators with citations from a multidisciplinary perspective". *http://arxiv.org/abs/1401.4321*

**Cronin, B.; Weaver, S.** (1995). "The praxis of acknowledgement: From bibliometrics to Influmetrics". *Revista Española de Documentación Científica,* vol. 18, n. 2, pp. 172-177. *http://dx.doi.org/10.3989/redc.1995.v18.i2.654*

**Davis, P.** (2013). "Visualizing article performance - Altmetric searches for appropriate display". *The Scholarly Kitchen*. *http://scholarlykitchen.sspnet.org/2013/09/30/visualizing-article-performance-altmetrics-searches-for-appropriate-display/*

**Jacso, P.** (2009). "Testing the Calculation of a Realistic h-index in Google Scholar, Scopus, and Web of Science for F.W. Lancaster". *Library Trends*, vol. 56, n. 4, pp. 784–815.

**Haustein, S.; Peters, I.; Bar-Ilan, J.; Priem, J.; Shema, H.; Tersliener, J.** (2014). "Coverage and adoption of altmetrics in the bibliometric community". *Scientometrics*. *http://dx.doi.org/10.1007/s11192-013-1221-3*

**Haustein, S.; Peters, I.; Sugimoto, C.; Thelwall, M.; Larivière, V.** (2013). "Tweeting biomedicine: An analysis of tweets and citations in the biomedical literature". *Journal of the American Society for Information Science and Technology. http://dx.doi.org/10.1002/asi.23101*

**Kurz, M.J.; Bollen, J.** (2010). "Usage bibliometrics". *Annual Review of Information Science and Technology,* vol. 44, pp. 1-64. *http://dx.doi.org/10.1002/aris.2010.1440440108*

**Li, X.; Thelwall, M.** (2012). "F1000 , Mendeley and Traditional Bibliometric Indicators". In *17th International Conference on Science and Technology Indicators*, vol. 3, pp. 1–11. *http://2012.sticonference.org/Proceedings/vol2/Li_F1000_541.pdf*